\documentclass[sigconf,nonacm]{acmart}

\setcopyright{acmlicensed}
\copyrightyear{2024}
\acmYear{2024}
\acmDOI{XXXXXXX.XXXXXXX}

\usepackage{microtype}
\usepackage{amsmath}
\usepackage{amsfonts}
\usepackage{placeins}
\usepackage{graphicx}
\usepackage{subcaption}
\usepackage{booktabs} % for professional tables
\usepackage{algorithm}
\usepackage{algpseudocode}
\usepackage{xspace}

\usepackage{hyperref}

\newcommand{\method}{\textsc{FastSample}\xspace}
\algnewcommand{\LineComment}[1]{\State \(\triangleright\) #1}

\begin{document}

\title{FastSample: Accelerating Distributed Graph Neural Network Training for Billion-Scale Graphs}


\author{Hesham Mostafa}
\affiliation{%
  \institution{Intel Corporation} 
  \city{}
  \country{}
}

\author{Adam Grabowski}
\affiliation{%
  \institution{Intel Corporation} 
  \city{}
  \country{}
}

\author{Md Asadullah Turja}
\affiliation{%
  \institution{UNC Chapel Hill} 
  \city{}
  \country{}
}

\author{Juan Cervino}
\affiliation{%
  \institution{ University of Pennsylvania} 
  \city{}
  \country{}
}

\author{Alejandro Ribeiro}
\affiliation{%
  \institution{ University of Pennsylvania} 
  \city{}
  \country{}
}

\author{Nageen Himayat}
\affiliation{%
  \institution{Intel Corporation} 
  \city{}
  \country{}
}

\renewcommand{\shortauthors}{Mostafa et al.}

\begin{abstract}
  Training Graph Neural Networks(GNNs) on a large monolithic graph presents unique challenges as the graph cannot fit within a single machine and it cannot be decomposed into smaller disconnected components. Distributed sampling-based training distributes the graph across multiple machines and trains the GNN on small parts of the graph that are randomly sampled every training iteration. We show that in a distributed environment, the sampling overhead is a significant component of the training time for large-scale graphs. We propose \method which is composed of two synergistic techniques that greatly reduce the distributed sampling time: 1)~a new graph partitioning method that eliminates most of the communication rounds in distributed sampling , 2)~a novel highly optimized sampling kernel that reduces memory movement during sampling. We test \method on large-scale graph benchmarks and show that \method speeds up distributed sampling-based GNN training by up to 2x with no loss in accuracy.

\end{abstract}



%% \received{20 February 2007}
%% \received[revised]{12 March 2009}
%% \received[accepted]{5 June 2009}

\maketitle
\section{Introduction}
Graphs are the natural representation for many types of real-world data. Recommendation graphs~\cite{ying2018graph}, knowledge graphs~\cite{bollacker2008freebase}, and molecular graphs~\cite{tran2023open} are few examples of graph-structured data where we need to extract high-level semantic information about the graph nodes, the graph links, or the whole graph. Graph Neural Networks (GNNs) have emerged as a highly-accurate deep-learning based solution for solving graph-related tasks~\cite{bronstein2017geometric,zhou2020graph,wu2020comprehensive}. The nature of graph data, however,  introduces many unique challenges when training GNNs on large graphs. Large graphs with millions of nodes and billions of edges, together with their node and edge features, cannot typically fit within the memory of a single machine. Distributed training is thus the natural approach to handle graphs of this size~\cite{Zheng_etal20,zheng2022distributed,mostafa2022sequential}. In  domains such as computer vision or natural language processing, it is often straightforward to split the data across several machines and train in a distributed data-parallel manner. The same cannot be easily done for graph data. The interconnected nature of the graph results in data dependencies between different machines no matter how we partition the graph~\cite{Karypis_etal97}. Therefore, in addition to parameter gradients, graph data also has to be periodically communicated between the different machines in each training iteration~\cite{Tripathy_etal20}.

Distributed GNN training falls into two main categories: full-graph (or full-batch) training~\cite{mostafa2022sequential,Md_etal21,wan2023scalable}, and sampling-based (or mini-batch) training~\cite{Zheng_etal20,zheng2022distributed}. In full-graph training, the GNN is trained on the entire distributed graph at once, while in sampling-based training, random subgraphs of the full graph are sampled each training iteration and the GNN is trained on these graph samples. While sampling-based training has many advantages such as a reduced memory footprint that makes it more practical for larger graphs~\cite{yang2022wholegraph}, and faster convergence due to the multiple weight updates per epoch~\cite{zeng2019graphsaint,Chiang_etal19}, the graph sampling operation can often consume a significant portion of the training time~\cite{zhang2021efficient}. Unlike mini-batching of i.i.d (Independent and identically distributed) data sources such as images or segments of text, graph mini-batching is computationally intensive as graph sampling involves traversing the graph to construct a random sub-graph~\cite{liu2021sampling}. We will show that the graph sampling overhead is significant in both single-node and distributed GNN training scenarios. The sampling overhead is more pronounced when working with distributed graphs as the sampling routines in each worker would typically need to traverse the graph across the graph partition boundaries. Distributed graph sampling thus requires several communication rounds to exchange vertex neighborhood information between the different machines hosting the graph partitions, which can slow down the sampling operation~\cite{Zheng_etal20,zheng2022distributed}.

In this paper, we present two synergistic techniques to address the issues outlined above and obtain dramatic improvements in time to convergence when running distributed sampling-based GNN training. Our contributions can be summarized as follows:
\begin{itemize}
  \item We present a highly optimized fused graph sampling kernel that is up to 2x faster than the highly efficient kernels in the popular Deep Graph Library (DGL)GNN package~\cite{Wang_etal19}.
  \item We observe that in many practical large-scale graphs, the size of the graph topology (the adjacency matrix) is often minuscule compared to the size of the graph's  node features. Motivated by that, we describe a hybrid partitioning scheme that only partitions the node features across the training machines. This leads to a significant reduction in the number of communication rounds in distributed sampling-based training, boosting performance by up to 1.5x.
\end{itemize}

We built a complete software library, \method, that implements these two techniques. We built the library on top of DGL. The ideas in \method, however, can be used to accelerate any other GNN framework.

\section{Related Work}
GNN training is typically done using open-source libraries such as DGL~\cite{Wang_etal19} or PyG~\cite{fey1903fast}. There has been several efforts that build on top of these libraries to enable scalable distributed GNN training: DistDGL~\cite{Zheng_etal20} builds on top of DGL by introducing a distributed sampling-based training pipeline backed up by a communication backend that uses RPC(Remote Procedure Calls). DistDGLv2~\cite{zheng2022distributed} improves on DistDGL using better load balancing and mini-batch generation heuristics. Quiver~\cite{tan2023quiver} scales up PyG model inference to multi-GPU settings and  uses dynamic heuristics to decide tensor placements and task allocations between the CPU and GPU. GraphLearn (formerly known as Aligraph)~\cite{zhu2019aligraph} is a high-performance distributed training system that uses smart neighbor caching to reduce communication in the distributed sampling step.

An extensive body of work looks at accelerating sampling-based GNN training in shared memory systems. These approaches, such as WholeGraph~\cite{yang2022wholegraph}, Zero-copy~\cite{min2021large}, and NeuGraph~\cite{Ma_etal19}, however, cannot scale to graphs that cannot realistically fit within the main memory of a single machine. A related approach is to replicate the full graph data (topology and features) across all machines and only synchronize gradients in each iteration~\cite{kaler2022accelerating}. This approach also does not scale. A work that bears some similarities to our hybrid partitioning scheme is P3~\cite{gandhi2021p3} which splits the node feature between machines along the feature dimension. Our scheme, instead, splits along the node dimension and ensures that each machine stores the node features of the seed nodes allocated to it. 

Accelerating graph sampling is an important problem. Several approaches have been proposed to bias the random neighborhood sampling operator to encourage node reuse during sampling~\cite{zheng2022distributed,Chiang_etal19}. Some Hybrid CPU-GPU approaches use two-step sampling where a large number of nodes is sampled on CPUs, which are then further sampled on GPUs to produce the graph minibatch~\cite{dong2021global,ramezani2020gcn}. These approaches, however, do not optimize the core sampling operation itself like what we do in the present work using our fused sampling kernel.

\section{Methods}
\subsection{Preliminaries}
Let $G(\mathcal{V},\mathcal{E})$ be a graph where $\mathcal{V}$ is the set of nodes and $\mathcal{E}\subset \mathcal{V}\times\mathcal{V}$ the set of edges. The input node features of node $i$ are $h_i^0$ and the feature vector of the edge from node $j$ to node $i$ is $e_{ji}$. Let $h_i^l$ be the feature vector of node $i$ at layer $l$, the GNN layer equations for $l=1,\ldots,L$, where $L$ is the number of GNN layers, are given by:
\begin{align}\label{eqn:Aggregate}
  m_{j\rightarrow i}^l = f_{message}(h_i^{l-1},h_j^{l-1},e_{ji};\theta^l)\\
  h_i^l = f_{node}(h_i^{l-1},Agg(\{m_{j\rightarrow i}^l : j \in \mathcal{N}(i)\});\phi^l),
\end{align}
where $f_{message}$ and $f_{node}$ are general functions whose learnable parameters at layer $l$ are $\theta^l$ and $\phi^l$, respectively. $\mathcal{N}(i)$ is the neighborhood of node $i$ in the graph, and $m_{j\rightarrow i}$ are messages from node $j$ to an adjacent node $i$. $Agg$ is a permutation-invariant aggregation operator such as $sum$ or $mean$. In node classification tasks, a subset of the nodes, $\mathcal{V}^C \subset \mathcal{V}$, is labeled and the labels are given during training. The goal of training is to minimize the prediction error on these labeled nodes. More precisely, for an $L$-layer GNN, the goal of training is to minimize the empirical loss:
\begin{equation}\label{eqn:ERM}
\mathcal{L} = \sum\limits_{v \in \mathcal{V}^C} \ell\left(h^L_v,y_v)\right),
\end{equation}
where $y_v$ is the label of node $v$ and $\ell$ is a standard classification loss such as the cross-entropy loss.

In an $L$-layer GNN, we need the $L$-hop neighborhood of node $v$ in order to produce the output at that node: $h^L_v$. For $L$ as low as 2 or 3, the sizes of these neighborhoods can become prohibitively large, a phenomenon known as the neighborhood explosion problem. Graph sampling is a standard approach to address this issue: instead of working with the full $L$-hop neighborhood, we sample a subset of the $L$-hop neighborhood and use this subsampled neighborhood during training. More precisely, for an $L$-layer GNN, we randomly pick a minibatch of training nodes ${\bf B}\equiv \mathcal{V}^L \subset \mathcal{V}^C$ and recursively sample their neighborhoods as follows for $l=L,\ldots,1$:
\begin{align}
  {\mathcal E}^{l-1} &= \bigcup\limits_{v \in \mathcal{V}^{l}} sample\_edges(v,N_l) \label{eq:sample_edges}     \\
  {\mathcal V}^{l-1} &= \{v : \exists u \text{ such that } e_{vu} \in {\mathcal E}^{l-1}\} \label{eq:get_nodes}
\end{align}
$sample\_edges(v,N_l)$ randomly samples $N_l$ incoming edges to node $v$. $N_l$ is an important hyper-parameter known as the sampling fanout. There are other more complex sampling operators~\cite{zou2019layer,zeng2019graphsaint,Chiang_etal19}. However, in this paper, we focus on the random neighborhood sampling operator outlined above as it is the one widely used in practice. 

After sampling, we construct $L$ bi-partite graphs, where the graph at layer $l$ is $\mathcal{G}^l=(\mathcal{V}^{l-1},\mathcal{V}^{l};\mathcal{E}^{l-1})$. The graph at layer $l$ has edges $\mathcal{E}^{l-1}$ from source nodes $\mathcal{V}^{l-1}$ to target nodes $\mathcal{V}^{l}$. These bi-partite graphs are also known as Message Flow Graphs (MFGs). GNN layer $l$ is applied to $\mathcal{G}^l$ and the output features for the nodes $\mathcal{V}^{l}$ are used as the input features of the source nodes of  $\mathcal{G}^{l+1}$

\subsection{Fused sampling}
\begin{figure*}[h]
  \centering
    \includegraphics[width = \textwidth]{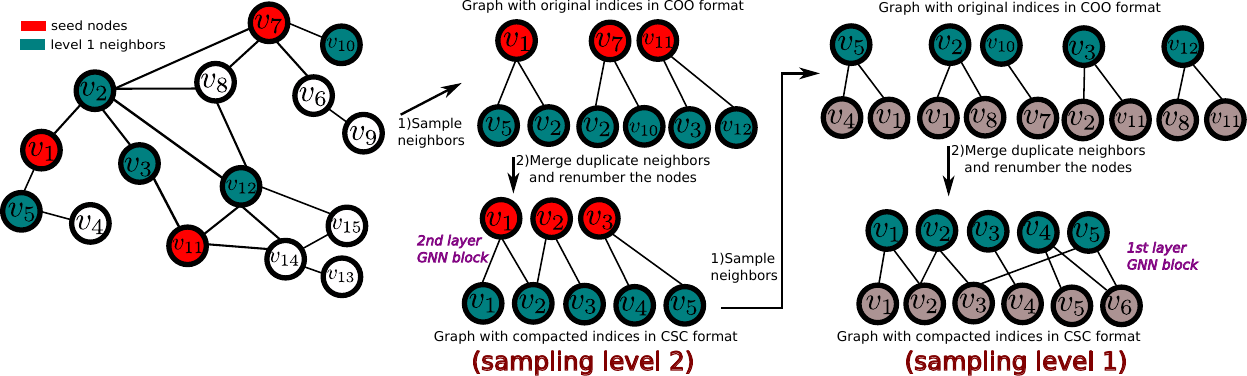} 
  \caption{Sampling the graph for a 2-layer GNN. The sampling fanout is $2$ throughout. The sampled neighbors  at one level become the seed nodes for the level below. The bi-partite graph produced at each level is used by the corresponding GNN layer. }
  \label{fig:sampling_1}
\end{figure*}

\begin{figure}[h]
  \centering
    \includegraphics[width = 0.5\textwidth]{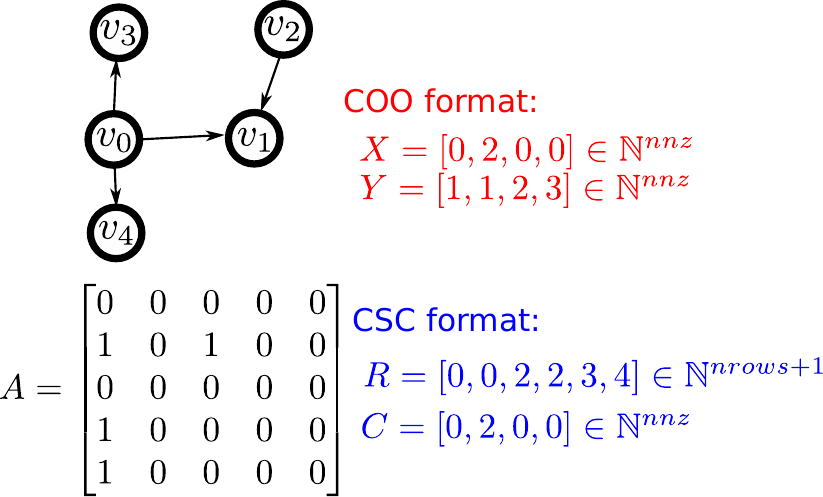} 
  \caption{Illustration of the COO and CSC representations of a sparse adjacency matrix $A$. $nnz$ is the number of non-zero elements. In COO, $(X[i],Y[i])$ are the row-column position of the $i^{th}$ non-zero element in $A$. In CSC, $R[k+1] - R[k]$ is the number of non-zero elements in row $k$ of $A$. The position of the non-zero elements in row $k$ are $C[R[k]:R[k+1]]$}
  \label{fig:coo_csr}
\end{figure}

The sampling operation has to be as efficient as possible as it is repeated every training iteration. Figure~\ref{fig:sampling_1} illustrates the steps involved in sampling the graph for a 2-layer GNN in the DGL library. As described in the previous section, the sampling operation has to be executed $L$ times for an $L$-layer GNN.  Sampling level $l$ involves two steps:
\begin{enumerate}
\item Sampling the neighbors of the seed nodes $\mathcal{V}^l$% (Eqs.~\ref{eq:sample_edges} and~\ref{eq:get_nodes})
\item Casting the resulting graph from step 1 as a bi-partite graph, $\mathcal{G}^l$, and reordering the node indices so that they are contiguous.
 \end{enumerate}
%The graph produced in step 1 is in COO (COOrdinate) format while the graph produced in step 2 is in CSC (Compressed Sparse Column) format. CSC is the preferred format for GNNs as it allows us to fetch a node's neighbors in O(1), i.e, independently of the size of the graph.
In DGL, both steps produce a graph in COO (COOrdinate) format. The preferred format for GNNs is CSC (Compressed Sparse Column) as it allows us to fetch a node's neighbors in O(1), i.e, independently of the size of the graph. We thus need an additional step to convert from the sampled graph from the COO to the CSC format. The two formats are illustrated in Fig.~\ref{fig:coo_csr}.

The conventional 2-step approach that is applied at each sampling level involves many redundant memory movements to write the output of step 1 in COO format to memory and then read it again to compact it and convert it to CSC format. Moreover, some information that is easily accessible in step 1 has to be re-computed in step 2 such as the number of sampled neighbors of each seed nodes (which can be less than the sampling fanout if the node has few neighbors).

A CSC matrix ${\bf A}$ is defined by the 2-tuple ${\bf A} \equiv (R,C)$ where $R$ and $C$ are the row pointer vector, and column indices vector, respectively. See Fig.~\ref{fig:coo_csr} for the definition of $R$ and $C$.  Algorithm 1 describes our fused sampling kernel which is applied at every sampling level to yield the CSC matrix for the bi-partite graph at that level as well as the seed nodes for the next level down. 
%The kernel avoids the overhead of creating the intermediate COO representation. 
Our kernel avoids the overhead of creating the intermediate COO representation.
It also constructs half of the CSC representation (the $R$ vector) practically for free during the actual sampling loop, and by sampling straight into the CSC format, it  avoids the expensive COO to CSC conversion. Our implementation is able to parallelize the two {\bf For} loops in the algorithm and the resulting kernel accelerates sampling by a large margin compared to vanilla DGL sampling.

\begin{algorithm}
  \caption{Fused sampling algorithm}
\begin{algorithmic}
\label{alg:fused}

\State {\bf Inputs}: Graph adjacency in CSC format: ${\bf A} \equiv (R_{G},C_{G})$
\State {\bf Inputs}: Unique seed nodes : $\mathcal{V}^l$
\State {\bf Inputs}: Sampling fanout : $N_l$
\State $R_l = zeros(|\mathcal{V}^l| + 1)$ \Comment{zero vector of size $|\mathcal{V}^l| + 1$}
\State $S_l = empty\_list$
\For  { $i := 0$ to $|\mathcal{V}^{l}|$} 
\State $v\leftarrow \mathcal{V}^l[i]$
\LineComment{Choose at most $N_l$ neighbors of $v$ }
\State $sampled \leftarrow Choose(C_{G}[R_{G}[v]:R_{G}[v+1]];N_l)$
\State $S_l.extend(sampled)$ 
\State  $R_l[i+1] \leftarrow R_l[i] + |Sampled|$
\EndFor

 \State $M = fill(|R_G|,-1)$ \Comment{vector of size $|R_G|$ filled with -1}
 \State $C_l = zeros(|S_l|)$
 \State $\mathcal{V}^{l-1} = empty\_list$
\State $idx = 0$
 \For  { $i := 0$ to $|S_l|$}
 \State $v\leftarrow S_l[i]$
 \If{$M[v] == -1$}
 \State $\mathcal{V}^{l-1}.append(v)$
 \State $M[v] \leftarrow idx$
 \State $idx \leftarrow idx+1$ 
 \EndIf
\State $C_l[i] \leftarrow M[v]$
\EndFor
\State  \Return $CSCMatrix(R_l,C_l),\mathcal{V}^{l-1}$

\end{algorithmic}
\end{algorithm}

\subsection{Distributed sampling and hybrid partitioning}
\begin{figure*}[h]
  \centering
    \includegraphics[width = \textwidth]{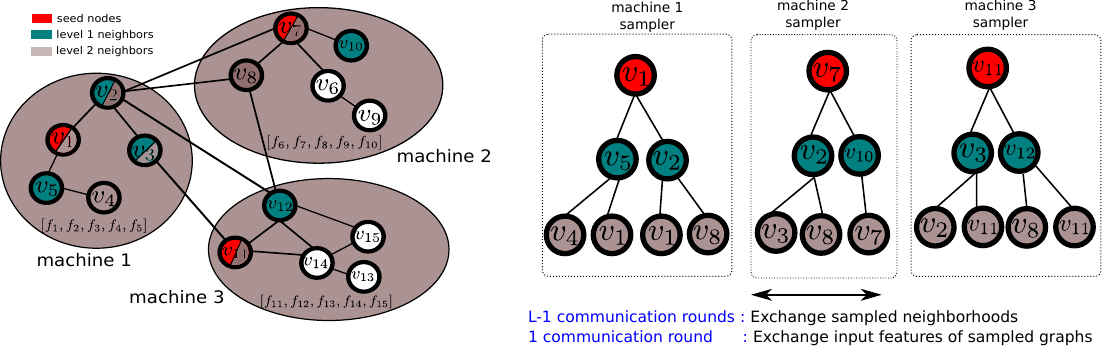} 
  \caption{Distributed sampling for a 2-layer GNN. The graph topology and node features are partitioned across three machines. At the top sampling level, each machine independently samples its local seed nodes and the neighbors of these seed nodes. For all sampling levels after that ($L-1$ levels), communication is needed. Communication is also needed to exchange the input node features of the sampled graphs. }
  \label{fig:sampling_dist}
\end{figure*}

In a distributed setting, we partition the graph across multiple machines. We consider edge-cut partitioning. Each worker stores one graph partition which comprises the partition node features as well as all incoming edges to the partition nodes from all other partitions. As illustrated in Fig.~\ref{fig:sampling_dist}, each machine independently samples seed nodes from the local partition, then independently samples the neighbors of these seed nodes. This can be done locally as each machine knows the neighbors of its local nodes (but not their features). All sampling levels below the top level, however, require machines to communicate. That is because each machine might need to sample the neighbors of nodes that are not in its local partition.

In general, for each sampling level after the top level, we would require two communication rounds, one round where machines submit their sampling requests to other machines, and one round where machines reply to each other with the the result of the sampling requests. For  $L$-level sampling, we thus need $2(L-1)$ communication rounds to sample the graph. Since the graph features are also partitioned, two additional communication rounds are needed at the end for machines to exchange the input features of the sampled graph, bringing the total number of communication rounds to $2L$. Note that each machine creates its own local sampled graph. The GNN is trained on the local sampled graph in each machine, and the parameter gradients are then synchronized.  

\begin{figure}[h]
  \centering
    \includegraphics[width = 0.5\textwidth]{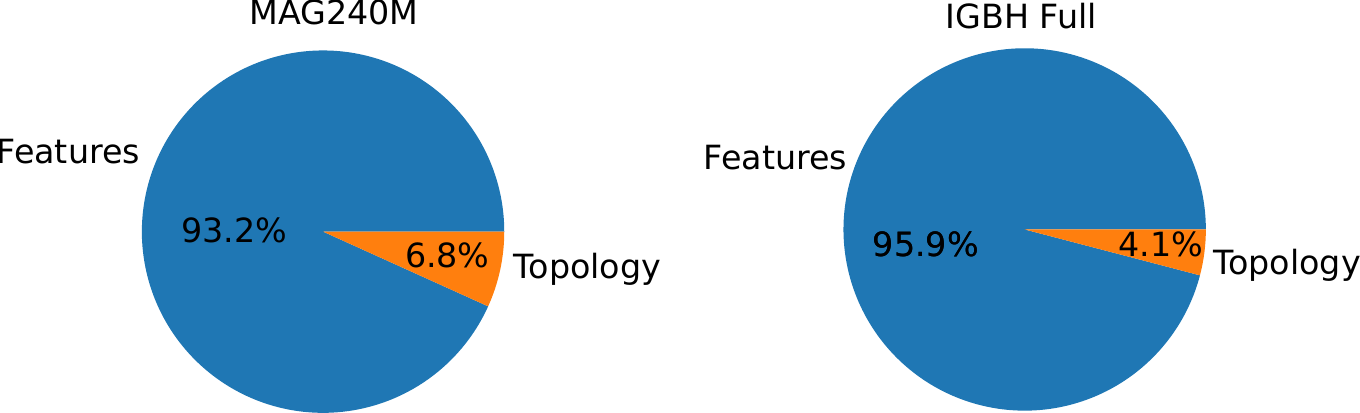} 
  \caption{Breakdown of the graph storage requirements for MAG240M~\cite{hu2021ogb} and IGBH-full~\cite{khatua2023igb}. Graph topology (adjacency matrix) is only a small fraction of total graph size}
  \label{fig:graph_breakdown}
\end{figure}

One important observation is that for many large-scale graphs, node features form the bulk of the graph size. This is illustrated in Fig.~\ref{fig:graph_breakdown} for two of the largest open-source graphs currently available. Motivated by this observation, we propose to duplicate the graph topology across all machines while partitioning the relatively much larger feature tensor. Sampling the graph topology can thus proceed independently in all machines. Communication is only needed to exchange the input node features of the sampled graph, bringing the number of communication rounds from $2L$ down to $2$. We call this scheme hybrid partitioning and we show that it significantly speeds up training while keeping memory consumption per machine within reasonable limits.

\section{Results}
We benchmark our contributions on two popular benchmark graphs: ogbn-products and ogbn-papers100M. The graph properties are outlined in Table~\ref{tab:dataset_properties}. We present results for different combinations of our techniques. We run our experiments on 2-socket machines, each equipped with two 4th Gen Intel Xeon Scalable Processors. All machines are connected by a 200Gbps Infiniband HDR fabric. In all experiments we use a 3-layer GraphSage model with hidden layer dimensions of 256. We use dropout between all layers. We use PyTorch 2.0.1 and DGL 1.1. We use FP32 numerical precision throughout. We build our \method library on top of  DGL.

In distributed training experiments,  we use torch\_ccl~\cite{torch_ccl} as our communication backend. torch\_ccl is the PyTorch wrapper for Intel's OneCCL collective communication library~\cite{oneccl}. Unlike popular distributed GNN training libraries such as DistDGL~\cite{Zheng_etal20}, we do not use point-to-point communication, but exclusively use synchronous collective communication calls such as $all\_to\_all$ and $all\_reduce$.  We use the metis library~\cite{Karypis_etal97} to partition the graph. Metis minimizes the number of edges that cross the partition boundaries. It  balances the number of nodes and edges in each partition so that all partitions are roughly of the same size. We also assign roughly the same number of labeled nodes to each partition. Since the top level sampling seeds are drawn from the labeled nodes, equalizing the number of labeled nodes across machines ensures they all have roughly the same number of seeds and can generate the same number of graph samples during each training epoch (see  Fig.~\ref{fig:sampling_dist}). In distributed training experiments, we always use a batch size of 1000 per machine.  We use a learning rate of 0.006. 

\begin{table}[h]
\caption{Graph Datasets}
\label{tab:dataset_properties}
\begin{center}
\begin{tabular}{l|cc}
\hline
  & \begin{tabular}{@{}c@{}}ogbn- \\ products\end{tabular} & \begin{tabular}{@{}c@{}}ogbn- \\ papers100M\end{tabular}  \\
    \hline
    \hline
    \# nodes & 2.5M & 111M  \\
    \# edges & 124M & 3.2B  \\    
    \# input features & 100 & 128  \\
    \# classes & 47 & 172 
\end{tabular}
\end{center}
\end{table}

\subsection{Fused sampling kernel}
\begin{figure}[h]
  \centering
  \begin{subfigure}{0.5\textwidth}
    \includegraphics[width = \textwidth]{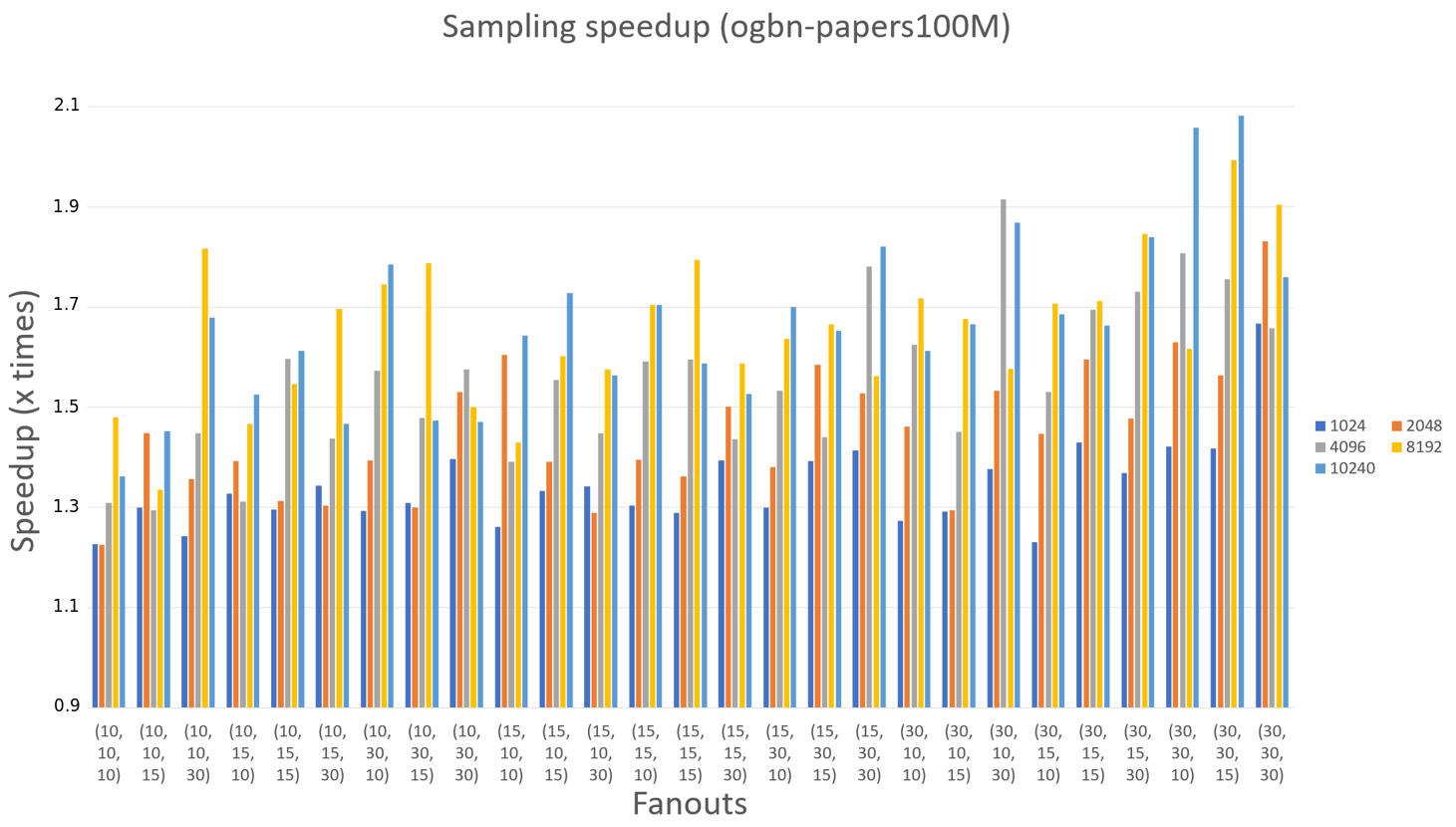} 
    \label{fig:speedup_a}
  \end{subfigure}
  \\
  \begin{subfigure}{0.5\textwidth}
    \includegraphics[width = \textwidth]{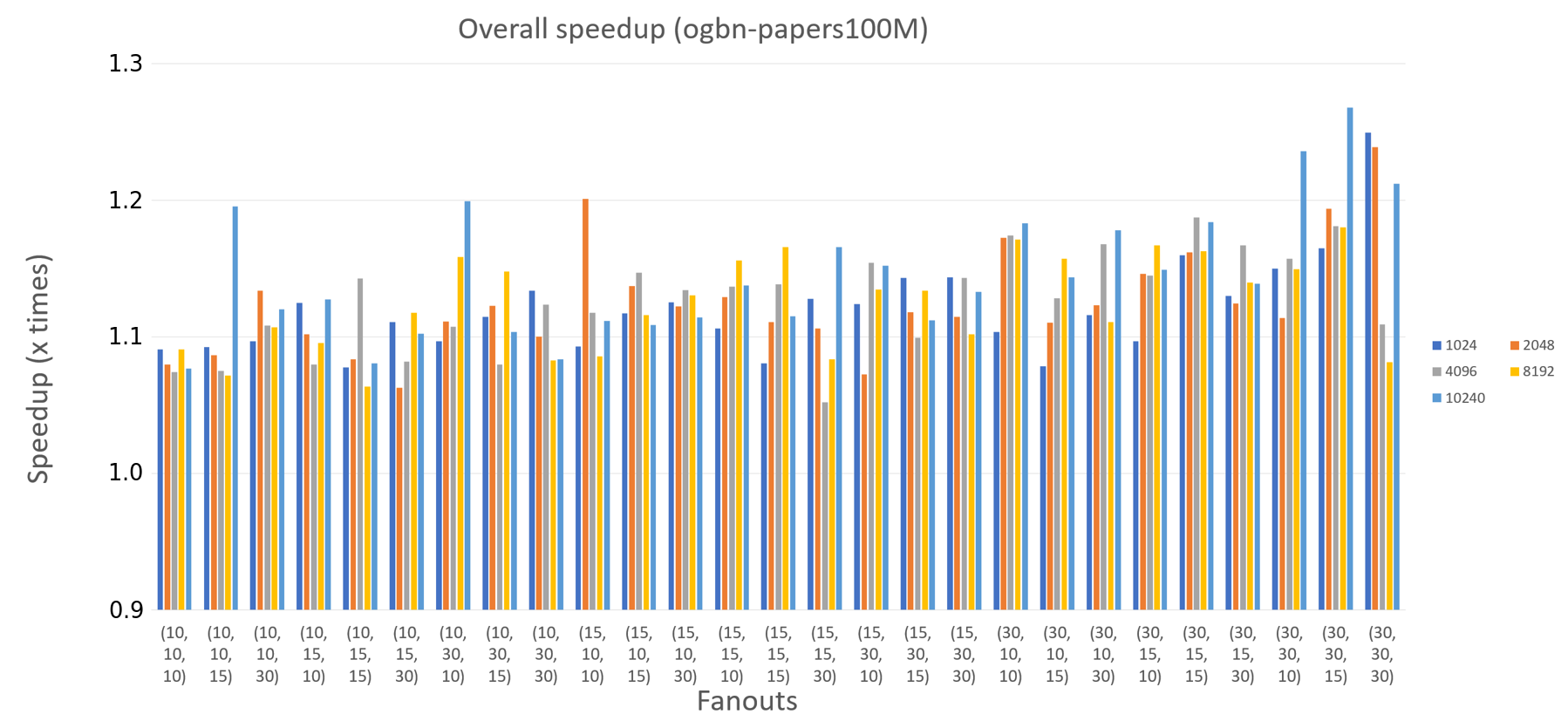} 
    \label{fig:speedup_b}    
  \end{subfigure}

%%   \centering
%%   \includegraphics[width=0.5\textwidth]{speedup_sampling_fused} 
  \caption{Speedup of the sampling time and the overall training time in single node training on ogbn-papers100M. Our baseline is the highly optimized sampling kernels in DGL. We plot the speedups for various mini-batch sizes ($1024,2048,\ldots,10240$) and different sampling fanout values for each of the three GNN layers in the model. Top panel: Speedup of the sampling operation. We obtain up to 2x speedup. Bottom panel: Speedup of overall training time (sampling + GNN training). We obtain speedups that are typically in the range of $10\%$ to $25\%$.} 
  \label{fig:fused_speedup}
\end{figure}

We evaluate the performance of our fused sampling kernel on ogbn-papers100M. There are two major hyper-parameters of the sampling operation: the batch size which is the number of top-level seed nodes, and the fan-out values at each sampling level. For a 3-layer GNN, the later can be written as a 3-tuple $(N_3,N_2,N_1)$ denoting the number of neighbors sampled at layers 3,2, and 1, respectively. As shown in Fig.~\ref{fig:fused_speedup}, our fused sampling kernel consistently speeds up the graph sampling operation across a wide range of batch sizes and sampling fanouts.  For the full training time (sampling + GNN training), we attain speedups of up to $25\%$.

\subsection{Hybrid partitioning}
\begin{figure}[h]
  \centering
  \begin{subfigure}{0.3\textwidth}
    \includegraphics[width = \textwidth]{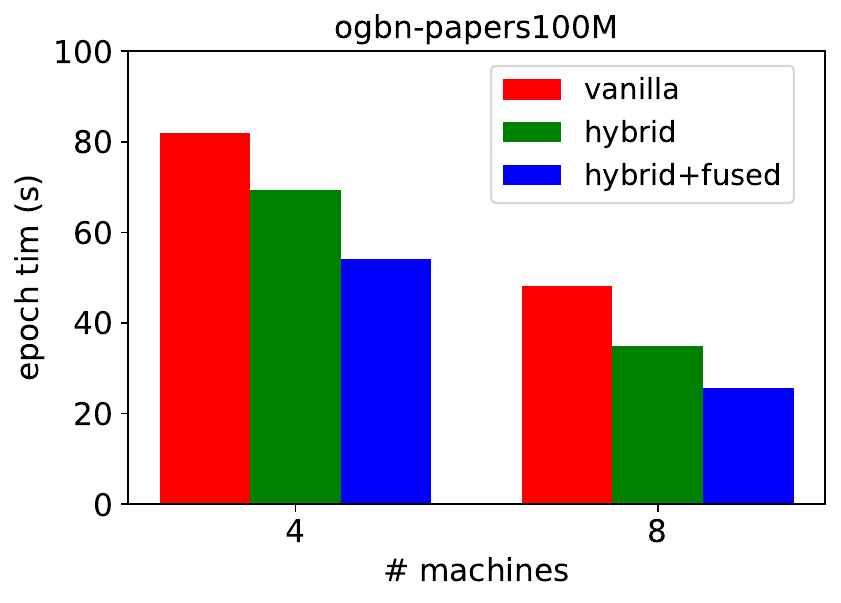} 
    \label{fig:hybrid_a}
  \end{subfigure}
  \\
  \begin{subfigure}{0.3\textwidth}
    \includegraphics[width = \textwidth]{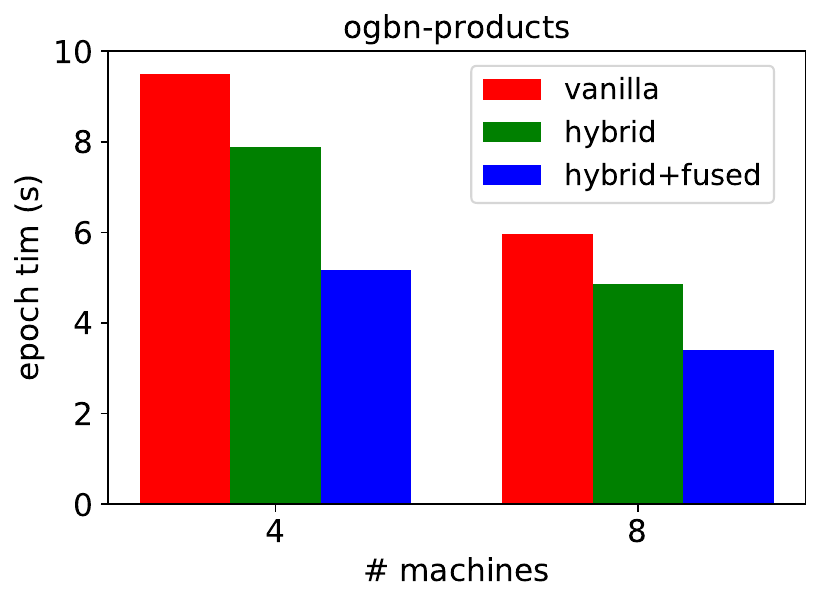} 
    \label{fig:hybrid_b}    
  \end{subfigure}
  
  \caption{Epoch times when training ogbn-products and ogbn-papers100M on 4 and 8 machines. We consider three scenarios: 1)vanilla training where we partition the graph topology and features and do not use fused sampling, 2)hybrid partitioning where we only partition the graph features and do not use fused sampling, 3)hybrid+fused where we only partition the graph features and use our fused sampling kernel.} 
  \label{fig:hybrid}
\end{figure}

To evaluate our hybrid partitioning method, we run distributed training  on 8 and 16 machines. Hybrid partitioning synergizes very well with our fused sampling kernel because the full graph topology is available on each machine and can thus be fed directly to our fused sampling kernel. Figure~\ref{fig:hybrid} shows the distributed training epoch times when training on ogbn-products  and ogbn-papers100M. Hybrid partitioning leads to a significant reduction in epoch time. When we combine hybrid partitioning with our fused sampling kernel, we get an even bigger performance boost. Hybrid partitioning together with fused sampling reduce the per-epoch training time by almost 2X when training ogbn-papers100M on 8 machines.

Note that hybrid partitioning and fused sampling have no effect on the convergence properties of the training. Activating or disabling these two techniques lead to mathematically equivalent training results. We also note that for fair comparison, all experiments use the same training and communication software infrastructure. This ensures that any improvements on the vanilla baseline are solely due to our new methods. 

%\subsection{Sampling with variable fanout}
%We evaluate the effectiveness of our variable fanout technique on ogbn-products and ogbn-papers100M. We vary the fanout values $(N^e_3, N^e_2, N^e_1)$ of a 3-layer GNN based on Table

% \begin{table}[h]
% \caption{Schedule for variable fanout}
% \label{tab:dataset_properties}
% \begin{center}
% \begin{tabular}{l|ccc}
% \hline
%   & \begin{tabular}{@{}c@{}}layer-3\end{tabular} & \begin{tabular}{@{}c@{}}layer-2\end{tabular}  & \begin{tabular}{@{}c@{}}layer-1\end{tabular}\\
%     \hline
%     \hline
%     \# 0-5 & 2.5M & 111M &  \\
%     \# edges & 124M & 3.2B &  \\    
%     \# input features & 100 & 128 &  \\
%     \# classes & 47 & 172 &
% \end{tabular}
% \end{center}
% \end{table}

%\FloatBarrier
\section{Conclusions}
Sampling-based training is becoming the method of choice to train GNNs on large graphs. As graph sizes continue to increase, distributed sampling-based training becomes the natural choice to accommodate massive graphs. In this paper, we have shown that the high-performance low-level CPU sampling kernels that are part of the DGL library, and that are in common use today, still leave plenty of performance on the table. Our fused kernel is able to consistently reduce the sampling time, in some cases accelerating the low-level graph sampling operations by up to 2X.

To make optimal use of our fused sampling kernel, we used a hybrid partitioning scheme that duplicates graph topology across all workers. The increased memory consumption per machine due to the duplicated adjacency matrix is an acceptable compromise since duplicating graph topology greatly reduces the number of communication rounds needed for sampling. The combination of hybrid partitioning and our fused sampling kernel is particularly attractive as it leaves the training iterations mathematically unchanged while greatly speeding up the distributed training (by up to 2X as shown in Fig.~\ref{fig:hybrid}).

In the future, our work can be extended in several ways. For example, we can combine our hybrid partitioning scheme with feature caching to cache frequently accessed remote node features in order to reduce communication volume. Or we can use an adaptive fanout schedule to dynamically adjust the sampling fanouts based on the training dynamics. 
%We believe there are plenty of optimization opportunities still left in the relatively nascent field of large-scale GNN training and deployment. 

% Convergence guarantee of this strategy
% Juan's part
% Write the convergence property of this new strategy

%\bibliographystyle{unsrtnat}
%\bibliography{biblio_small.bib}

%\printbibliography

\end{document}